\documentclass[12pt]{iopart}

\usepackage{graphicx}
\usepackage{amstext}
\begin{document}

\title[Superbunching pseudothermal light]{Superbunching pseudothermal light with intensity modulated laser light and rotating groundglass}

\author{Yu Zhou$^1$, Xuexing Zhang$^2$, Zhengpeng Wang$^2$, Feiye Zhang$^2$, Hui Chen$^2$, Huaibin Zheng$^2$, Jianbin Liu$^{2,*}$, Fu-li Li$^1$, Zhuo Xu$^2$}

\address{$^1$ MOE Key Laboratory for Nonequilibrium Synthesis and Modulation of Condensed Matter, Department of Applied Physics, Xi'an Jiaotong University, Xi'an 710049, China}
\address{$^2$ Electronic Materials Research Laboratory, Key Laboratory of the Ministry of Education \& International Center for Dielectric Research, School of Electronic and Information Engineering, Xi'an Jiaotong University, Xi'an 710049, China}

\ead{liujianbin@xjtu.edu.cn}
\vspace{10pt}
\begin{indented}
\item[]September 2018
\end{indented}

\begin{abstract}
Pseudothermal light by scattering laser light from rotating groundglass has been extensively employed to study optical coherence in both classical and quantum optics ever since its invention in 1960s.  In this paper, we will show that by replacing the invariant intensity laser light in pseudothermal light source with intensity modulated laser light, superbunching pseudothermal light can be obtained. Two-photon interference in Feynman's path integral theory is employed to interpret the phenomenon. Two-photon superbunching is experimentally observed by employing common instruments in optical laboratory. The proposed superbunching pseudothermal light is helpful to understand the physics of two-photon bunching and superbunching, and the difference between classical and quantum interpretations of the second- and higher-order interference of light.
\end{abstract}

\submitto{\NJP}

\section{Introduction}

Thermal light is the most common natural light, which has played important role in the development of optical coherence theory \cite{mandel-book}. For instance, sunlight was employed by Young to prove that light is wave instead of particle in his famous double-slit interference experiment \cite{young}, in which sunlight belongs to thermal light. Hanbury Brown and Twiss (HBT) observed that randomly emitted photons in thermal light tend to come in bunches rather than randomly \cite{HBT}, which is usually thought as the foundation of modern quantum optics \cite{glauber}. However, it is difficult to implement the HBT experiments with true thermal light due to its ultra short coherence time and low degeneracy factor \cite{mandel-book,HBT-2}. 

In 1964, Martienssen and Spiller invented a ``novel light source'' by passing a single-mode continuous-wave laser light beam through a rotating groundglass, which is latter called pseudothermal light source. The coherence properties of pseudothermal light are similar as the ones of true thermal light except its degeneracy factor can be much larger than 1 and its coherence time can be varied between $10^{-5}$ s to 1 s \cite{martienssen}. Since its invention, this new type of light has been widely employed in the experiments when thermal light is needed. For instance, the photon number statistical distribution emitted by Gaussian radiation source was first measured with pseudothermal light \cite{arecchi}. Pseudothermal light is also widely employed in ghost imaging experiments \cite{valencia,ferri,crosby}, multi-photon interference \cite{scarcelli,chen}, non-invasive imaging \cite{bertolotti,katz}, delayed-choice eraser \cite{peng} and so on.  The experiments with pseudothermal light is much simpler than the one with true thermal light while, the observed phenomenons remain similar \cite{valencia,wu-true}.

In pseudothermal light source, the intensity of the incident laser light before rotating groundglass is constant, which causes the photons in the scattered light are bunched as the ones in thermal light. The degree of second-order coherence, $g^{(2)}(0)$, of thermal light equals 2. How will the coherence properties of pseudothermal light change if the intensity of the incident laser light is not constant? In order to answer this interesting and important question, we will study the coherence properties, especially $g^{(2)}(0)$, of the pseudothemal light by varying the intensity of the incident laser light. In most cases, we find that $g^{(2)}(0)$ can exceed 2. Based on the definition of photon bunching and superbunching \cite{ficek}, we name this new type of light superbunching pseudothermal light \cite{zhou}. It seems to be a minor modification of the well-known pseudothermal light source. However, no one had done this for more than 50 years until recently \cite{zhou}. This new proposed superbunching pseudothermal light is expected to be helpful for the development of optical coherence theory, especially for understanding the physics of the second- and higher-order interference of light.

\section{Theory}

There are laser light and linear optical elements in the superbunching pseudothermal light source. Hence both classical and quantum theories can be employed to interpret the superbunching effect in our system \cite{klyshko}. In fact, the calculated results in these two different theories are equivalent \cite{glauber-2,sudarshan}, even though the physical interpretations are different. For instance, two-photon bunching of thermal light is interpreted by two-photon interference in quantum theory \cite{purcell,fano}, while the same phenomenon is interpreted by intensity fluctuations correlation in classical theory \cite{HBT-2}. However, the degree of second-order coherence of thermal light is calculated to be 2 in both quantum and classical theories \cite{mandel-book}. In our earlier studies, we have employed two-photon interference in Feynman's path integral theory to discuss the second-order interference of two independent light beams \cite{zhou,liu-1,liu-2,liu-3,liu-4,liu-5,liu-6}. It is found that the advantages the method is not only the straightforward calculation, but also easy to connect the mathematical calculations with the corresponding physical interpretations.  In this section, we will employed the same method to calculate the second-order coherence function of superbunching pseudothermal light, hoping to get a better understanding about the physics of two-photon bunching and superbunching.

\begin{figure}[htbp]
\centering
\includegraphics[width=60mm]{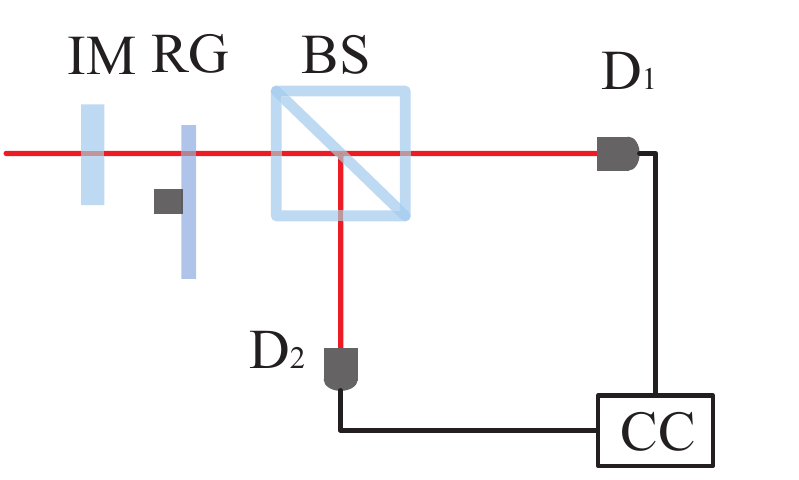}
\caption{Superbunching pseudothermal light measured by the HBT interferometer. IM: intensity modulator. RG: rotating groundglass. BS: 1:1 beam splitter. D: single-photon detector. CC: Two-photon coincidence counting system. The intensity of the single-mode continuous-wave laser light before IM is constant. The intensity variation is applied by IM. The second-order coherence function of the scattered light is measured by the HBT interferometer. The positions of D$_1$ and D$_2$ are symmetrical when the second-order temporal coherence function is measured.}\label{calculation}
\end{figure}

The scheme for calculating the second-order coherence function of superbunching pseudothermal light is shown in Fig. \ref{calculation}. The intensity of the incident laser light before intensity modulator (IM) is constant. After passing IM, the laser light with modulated intensity is scattered by the rotating groundglass (RG). A 1:1 beam splitter (BS), two single-photon detectors (D$_1$ and D$_2$), and two-photon coincidence counting system (CC) consist a standard HBT interferometer \cite{HBT}. There are two different ways for two photons, a and b, in superbunching pseudothermal light to trigger a two-photon coincidence count event at D$_1$ and D$_2$ in the scheme shown in Fig. \ref{calculation} \cite{zhou,fano,shih-book}. One way is that photon a is detected by D$_1$ and photon b is detected by D$_2$. The other way is that photon a is detected by D$_2$ and photon b is detected by D$_1$. If these two different ways are in principle indistinguishable, the second-order coherence function in Fig. \ref{calculation} is \cite{liu-6,feynman}
\begin{eqnarray}\label{G2-1}
&&G^{(2)}(\vec{r}_1,t_1;\vec{r}_2,t_2)=
\langle |P_{a1}P_{b2}e^{i\varphi_a}A_{a1}e^{i\varphi_b}A_{b2}+P_{a2}P_{b1}e^{i\varphi_a}A_{a2}e^{i\varphi_b}A_{b1} |^2 \rangle,
\end{eqnarray}
where $(\vec{r}_1,t_1)$ and $(\vec{r}_2,t_2)$ are the space-time coordinates of the photon detection events at D$_1$ and D$_2$, respectively. $P_{\alpha \beta}$ and $A_{\alpha \beta}$ are the probability and probability amplitude for photon $\alpha$ is detected by D$_\beta$, respectively ($\alpha=$a and b, $\beta=$1 and 2). $\varphi_a$ and $\varphi_b$ are the initial phases of photons a and b, respectively. $A_{a1}A_{b2}$ is the two-photon probability amplitude corresponding to photon a is detected by D$_1$ and photon b is detected by D$_2$. $A_{a2}A_{b1}$ is defined similarly. $\langle ... \rangle$ is ensemble average, which can be treated as time average for ergodic system \cite{mandel-book}. When the intensity of the incident laser light before RG is constant, all the four probabilities, $P_{a1}$, $P_{a2}$, $P_{b1}$, and $P_{b2}$, are equal. Equation (\ref{G2-1}) can be simplified as $\langle |A_{a1}A_{b2}+A_{a2}A_{b1} |^2 \rangle$, which is identical to the quantum interpretation of two-photon bunching of thermal or pseudothermal light \cite{fano,shih-book}. 

In order to simplify the calculation, D$_1$ and D$_2$ are assumed to be in the symmetrical positions, \textit{i.e.}, the distance between BS and D$_1$ equals the one between BS and D$_2$ and the transverse positions of both detectors are symmetrical. With the above simplifications, Eq. (\ref{G2-1}) can be simplified as
\begin{eqnarray}\label{G2-2}
G^{(2)}(t_1,t_2)=\langle |\sqrt{P_{a1}P_{b2}}A_{a1}A_{b2}+\sqrt{P_{a2}P_{b1}}A_{a2}A_{b1} |^2 \rangle
\end{eqnarray}
by dropping the space coordinates. If the employed laser light is assumed to be monochromatic and pseudothermal light source is assumed to be a point light source, the probability amplitude, $A_{\alpha\beta}$, can be expressed as \cite{QFT}
\begin{eqnarray}\label{G2-3}
A_{\alpha\beta}=\frac{e^{-i{\omega (t_\beta-t_\alpha)}}}{r_{\alpha\beta}},
\end{eqnarray}
in which $t_{\beta}$ is the detection time at D$_\beta$, $t_\alpha$ is the time of photon $\alpha$ in the interferometer, and $r_{\alpha\beta}$ is the distance between the photon at $t_{\alpha}$ and $t_{\beta}$. Substituting Eq. (\ref{G2-3}) into Eq. (\ref{G2-2}), the second-order temporal coherence function can be expressed as \cite{zhou,liu-6,shih-book}
\begin{eqnarray}\label{G2-4}
&&G^{(2)}(t_1,t_2)\nonumber\\
&\propto& \langle P_{a1}P_{b2} \rangle + \langle P_{a2}P_{b1}\rangle+2\langle \sqrt{P_{a1}P_{b2}P_{a2}P_{b1}} \rangle \text{sinc}^2\frac{\Delta \omega (t_1-t_2)}{2},
\end{eqnarray}
in which $\Delta \omega$ is the frequency bandwidth of pseudothermal light caused by rotating groundglass. If we assume the probability, $P_{\alpha\beta}$, is proportional to the intensity of the incident laser light at $t_{\alpha\beta}$, we have $P_{\alpha\beta} \propto I(t_\beta-\tau_\alpha)$, where $t_{\alpha\beta}$ equals $t_\beta-\tau_\alpha$ and $\tau_\alpha$ is the traveling time of photon $\alpha$ goes to D$_\beta$. Since D$_1$ and D$_2$ are in the symmetrical positions, the traveling time of photon $\alpha$ goes to D$_1$ equals the one of photon $\alpha$ goes to D$_2$. If stationary light is employed, Eq. (\ref{G2-4}) can be simplified as \cite{mandel-book}
\begin{eqnarray}\label{G2-5}
G^{(2)}(t_1-t_2)\propto \Gamma(t_1-t_2) [1+ \text{sinc}^2\frac{\Delta \omega (t_1-t_2)}{2}],
\end{eqnarray}
where $\Gamma(t_1-t_2)$ is defined as $ \frac{\langle I(t_1-\tau_a) I(t_2-\tau_b)  \rangle}{\langle I(t_1-\tau_a) \rangle \langle I(t_2-\tau_b)  \rangle}$ and  the ensemble average is calculated by summing the traveling time $\tau_a$ and $\tau_b$ from $\tau_0$ to infinity. $\tau_0$ is the traveling time of photon between RG and D$_1$ or D$_2$. $\Gamma(t_2-t_1)$  equals  $\Gamma(t_1-t_2)$ for stationary light. One important thing worthy of noticing is that $\Gamma(t_1-t_2)$ describes the intensity correlation of laser light before the RG, not the correlation of light at D$_1$ and D$_2$. When the intensity of laser light before RG is constant, $\Gamma(t_1-t_2)$ equals 1 and Eq. (\ref{G2-5}) becomes the second-order temporal coherence function of thermal or pseudothermal light \cite{martienssen,shih-book}. 

Let us take two types of modulated intensities for example to calculate the second-order coherence function of superbunching pseudothermal light. One is sinusoidal wave. The intensity at time $t_\beta-\tau_\alpha$ is
\begin{eqnarray}\label{I-s}
I(t_\beta-\tau_\alpha)=I_0[1+C\cos\omega_0(t_\beta-\tau_\alpha)] ,
\end{eqnarray}
where $I_0$ is a constant intensity,  $C$ is a constant and in the regime of $[0,1]$, and $\omega_0$ is the frequency of sinusoidal wave. Substituting Eq. (\ref{I-s}) and $P_{\alpha\beta} \propto I(t_\beta-\tau_\alpha)$ into Eq. (\ref{G2-4}), it is lengthy but straightforward to have the second-order temporal coherence function of superbunching psedudothermal light with sinusoidal intensity laser light as the input,
\begin{eqnarray}\label{G2-s}
G^{(2)}_s(t_1-t_2)\propto [1+2C\cos^2\frac{\omega_0(t_1-t_2)}{2}] [1+ \text{sinc}^2\frac{\Delta \omega (t_1-t_2)}{2}].
\end{eqnarray}
When $C$ equals 1, it corresponds to the intensity of the modulated laser light beam is a perfect sinusoidal wave. The maximal value of $G^{(2)}_s(t_1-t_2)$ in this case equals 6. The background value of $G^{(2)}_s(t_1-t_2)$ is 2 instead of 1. Hence the degree of second-order coherence equals 3, which is larger than 2.

The other one is that the intensity of the modulated laser light follows the distribution of white noise. Based on the Wiener-Khinchin theorem, the auto-correlation function of the electric field can be expressed as \cite{goodman-s}
\begin{eqnarray}\label{I-r}
\gamma(t_1-t_2)=\int_{-\infty}^{\infty}f(\nu)e^{-i2\pi\nu t}d\nu,
\end{eqnarray}
in which $f(\nu)$ is the power spectral density and can be treated as a constant within the frequency regime of white noise. For Gaussian variates, the correlation of intensity can be expressed by the correlation function of electric field \cite{mandel-book},
\begin{eqnarray}\label{I-r1}
\Gamma(t_1-t_2)=1+|\gamma(t_1-t_2)|^2.
\end{eqnarray}
If the frequency of white noise is $\nu_0$, substituting Eqs. (\ref{I-r}) and (\ref{I-r1}) into Eq. (\ref{G2-5}), it is straightforward to have the second-order temporal coherence function of superbunching pseudothermal light with random intensity laser light as the input,
\begin{eqnarray}\label{G2-r}
G^{(2)}_r(t_1-t_2)\propto [1+\text{sinc}^2 \pi \nu_0(t_1-t_2)] [1+ \text{sinc}^2\frac{\Delta \omega (t_1-t_2)}{2}].
\end{eqnarray}
The maximal and background values of $G^{(2)}_r(t_1-t_2)$ in Eq. (\ref{G2-r}) are 4 and 1, respectively. Hence the degree of second-order coherence equals 4, which is also larger than 2.

\section{Experiments}

\begin{figure}[htbp]
\centering
\includegraphics[width=91mm]{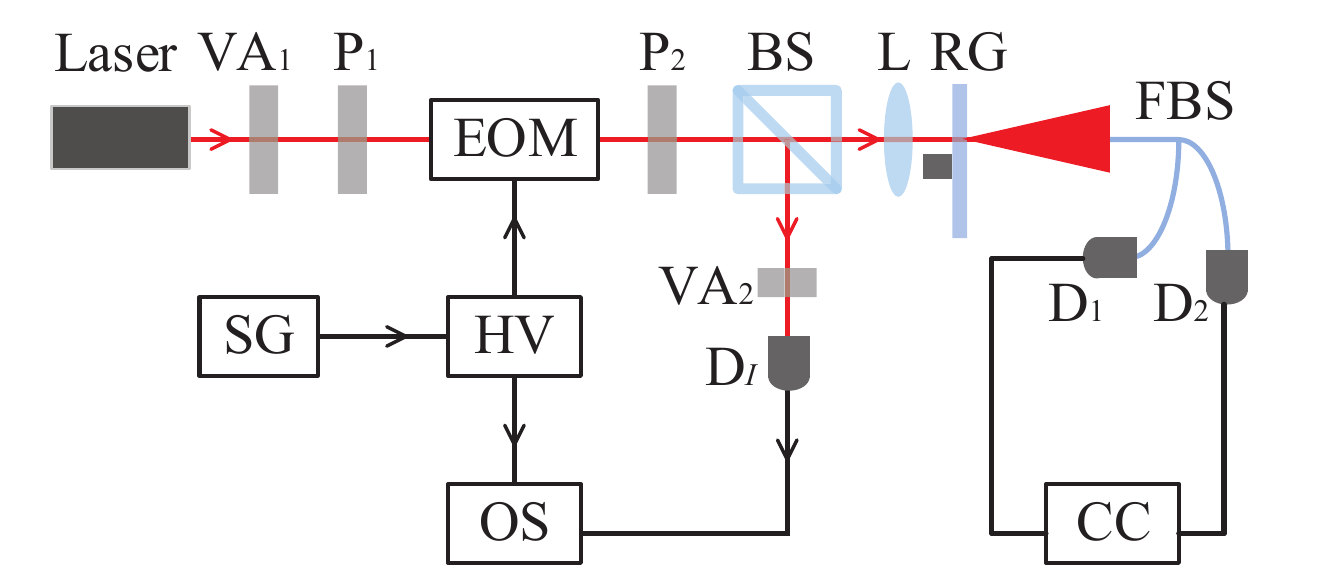}
\caption{Scheme for superbunching pseudothermal light with modulated laser light beam. Laser: single-mode continuous-wave laser. VA: variable attenuator. P: polarier. SG: signal generator. EOM: electro-optical modulator. HV: high-voltage amplifier. OS: oscilloscope. BS: 1:1 beam splitter. L: lens. RG: rotating groundglass. FBS: 1:1 fiber beam splitter. D$_I$: intensity detector. D$_1$ and D$_2$: single-photon detectors. CC: two-photon coincidence counting system. See text for detail explanations of the experimental setup.}\label{setup}
\end{figure}

In this section, we will employ the scheme shown in Fig. \ref{setup} to verify our theoretical predictions. The employed laser is a single-mode continuous-wave laser with center wavelength at 780 nm and frequency bandwidth of 200 kHz (Newfocus, SWL-7513). VA$_1$ and VA$_2$ are two variable attenuators to control the intensities of light beams. P$_1$ and P$_2$ are vertically and horizontally polarized linear polarizers, respectively. EOM is an electro-optical modulator, which is employed to change the polarization of the incident laser light according to the applied voltage via the high voltage amplifier (HV) (Thorlabs, EO-AM-NR-C1 and HVA200). Signal generator (SG) is employed to generate the voltage signal to control the intensity of light together with P$_1$, EOM, and P$_2$. The amplified voltage signal from HV is employed to drive EOM and the monitoring output of HV is connected to an oscilloscope (OS) to monitor the applied voltage. The ratio between the amplified and monitoring voltages of HV is 20. A beam splitter (BS) is employed to divide the modulated laser light beam into two. One beam is detected by an intensity detector (D$_I$) to monitor the light intensity. The other beam is incident to a rotating groudglass (RG) to generating pseudothermal light \cite{martienssen}. FBS is a 1:1 fiber beam splitter. D$_1$ and D$_2$ are two single-photon detectors (PerkinElmer, SPCM-AQRH-14-FC). CC is a two-photon coincidence count detecting system (Becker\&Hickl GmbH, DPC230). FBS, D$_1$, D$_2$ and CC are employed to record the second-order temporal coherence function of the scattered light, which is equivalent to a HBT interferometer \cite{HBT}. 

\begin{figure}[htbp]
\centering
\includegraphics[width=80mm]{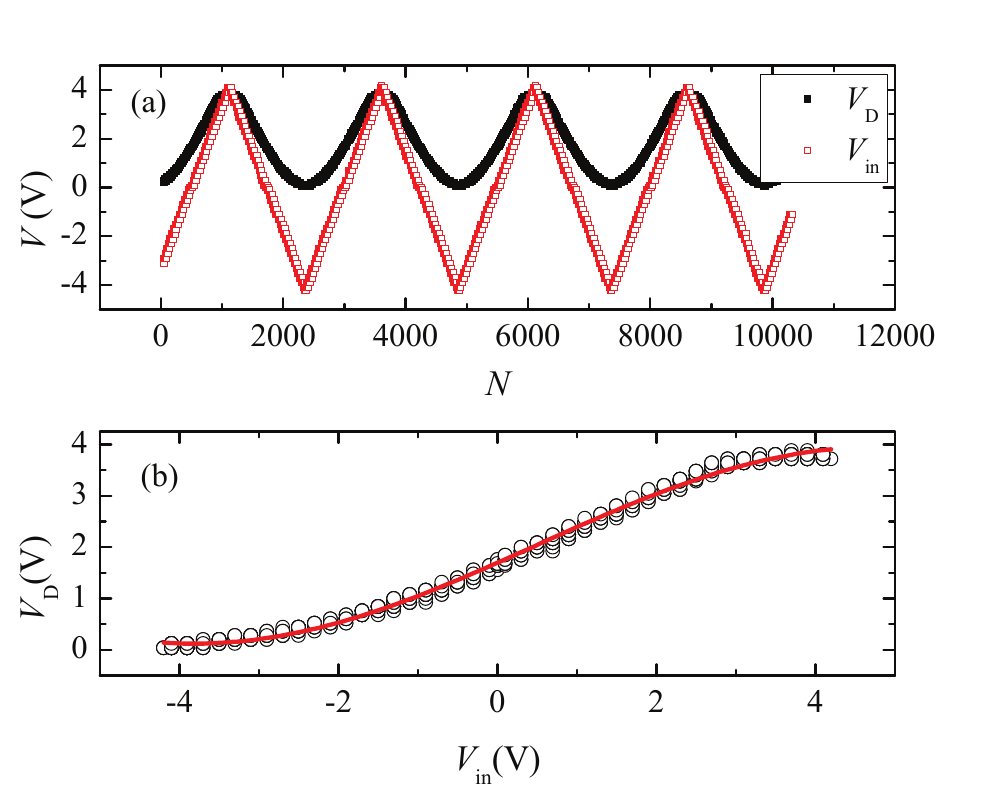}
\caption{Relation between the applied voltage of EOM and the intensity of the modulated laser light. Figure (a) is the output of the oscilloscope. $N$ is the number of time series recorded by OS. $V$ is the output voltage.  The red empty squares are the applied voltages of the EOM, which are generated by SG and written as $V_{\text{in}}$. The black squares are the output of D$_I$, which is written as $V_{\text{D}}$. Figure (b) is drawn by selecting the data when $N$ is between 2330 and 3660 in Fig. (a), in which the applied voltage increase from -4.1 V to 4.1 V. The red solid line is the sinusoidal fitting of the data. The intensity of the modulated light varies sinusoidally as the applied voltage increases linearly.}\label{ratio}
\end{figure}

We first measure the relationship between the applied voltage of EOM and the intensity of the modulated laser light, which is shown in Fig. \ref{ratio}. Figure \ref{ratio}(a) is the output of the oscilloscope. A triangular-wave voltage signal is applied on the EOM, which is shown by the red empty squares and written as $V_{\text{in}}$. $V_{\text{D}}$ is the output of the monitoring detector, D$_I$, which is proportional to the intensity of the modulated laser light and shown by black squares. As the applied voltage increases linearly, the intensity of the modulated laser light varies sinusoidally. In order to analyze the relationship between the applied voltage and intensity, we choose the data between 2330 and 3660 in Fig. \ref{ratio}(a) when the applied voltage increases from -4.1 V to 4.1 V. The ratio between the output and input voltages of HV is 20. Hence the voltage applied on the EOM increases from -80.4 V to 80.4 V. The minimum intensity of the modulated laser light is $0.04\beta$, where $\beta$ is a constant determined by the detector, oscilloscope, and other electronic elements. The maximum value of the modulated laser is $3.80\beta$, which is obtained when the applied voltage equals 4.1 V. The circles in Fig. \ref{ratio}(b) are the measured results and the red solid line is sinusoidal fitting of the measured results. The fitted equation is 
\begin{eqnarray}\label{fitted}
V_{\text{D}}=2.04+1.92\sin \frac{\pi \times (V_{\text{in}}-0.49)}{8.65}.
\end{eqnarray}
When the applied voltages are in the domain near 0 V, the intensities of the modulated laser light are linearly dependent on the applied voltages. As $V_{\text{in}}$ falls apart from 0 V, the output intensity varies sinusoidally, which is obvious from the measured results in Fig. \ref{ratio}(b). We will assume that the linear response regime is employed in our experiments.

\begin{figure}[htbp]
\centering
\includegraphics[width=95mm]{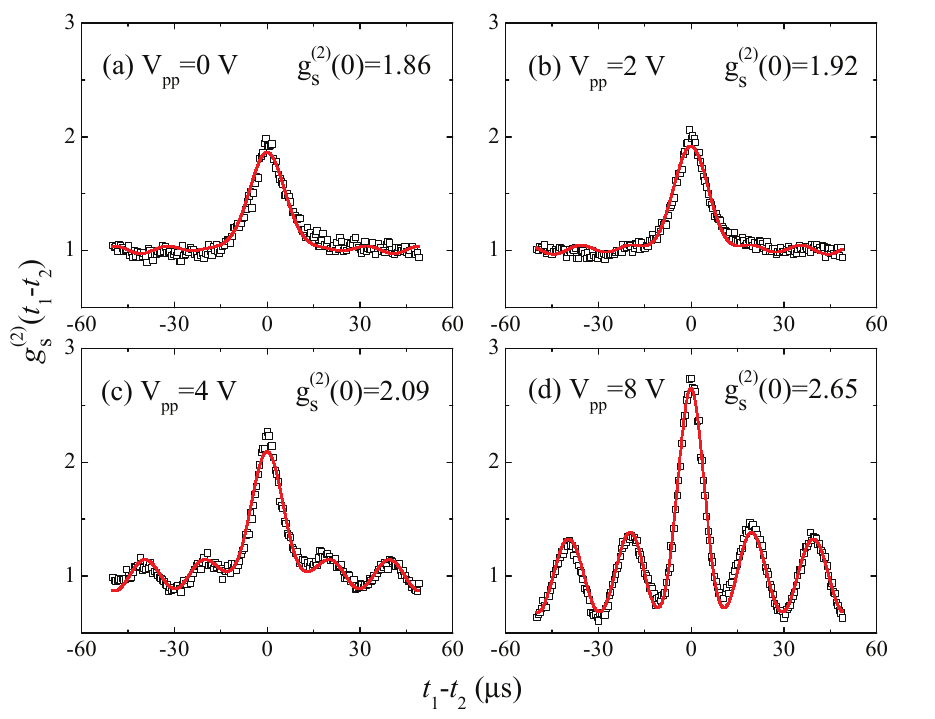}
\caption{The second-order temporal coherence function of superbunching pseudothermal light with sinusoidal voltage signal applied on EOM by varying $V_{\text{pp}}$. $V_{\text{pp}}$ is the peak-to-peak voltage of sinusoidal voltage signals.The frequency of the sinusoidal signal is 50 kHz. $t_1-t_2$ is the time difference between two photon detection events within a two-photon coincidence count. $g^{(2)}_\text{s}(t_1-t_2)$ is the normalized second-order temporal coherence function with sinusoidal voltage signals. The empty squares and red curves in (a)-(d) are measured results and theoretical fittings, respectively. All the conditions in (a) - (d) are the same except the applied voltages on the EOM are 0 V, 2 V, 4 V, and 8V, respectively.}\label{voltage}
\end{figure}

\begin{figure}[htbp]
\centering
\includegraphics[width=70mm]{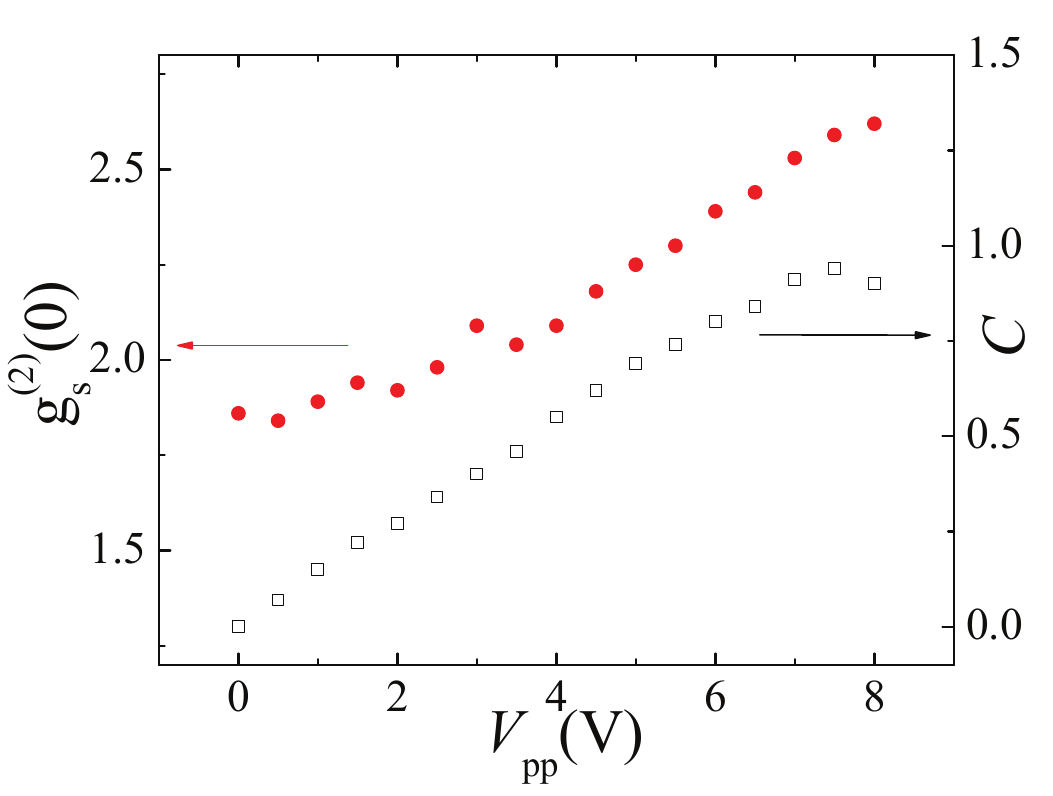}
\caption{The measured degree of second-order coherence and $C$ versus voltages when the frequency of sinusoidal signal is 50 kHz. $g^{(2)}_\text{s}(0)$ is the degree of second-order coherence and $C$ is the coefficient in Eq. (\ref{I-s}). The red circles are the measured $g^{(2)}_\text{s}(0)$, which is calculated by fitting the data with Eq. (\ref{G2-s}). The empty squares are the measured coefficient of the sinusoidal voltage signal. As $V_{\text{pp}}$ increases from 0 V to 8 V, the coefficient, $C$, increases from 0 to nearly 1 and $g^{(2)}_\text{s}(0)$ increases from less than 2 to larger than 2.}\label{g2-C}
\end{figure}

Figure \ref{voltage} shows the measured second-order coherence function of superbunching pseudothermal light when 50 kHz sinusoidal voltage signals are applied on the EOM. $g^{(2)}_\text{s}(t_1-t_2)$ is the normalized second-order temporal coherence function with sinusoidal voltage signals and $t_1-t_2$ is the time difference between two photon detection events within a two-photon coincidence count. The empty squares are measured results and red curves are theoretical fitting by employing Eq. (\ref{G2-s}). When $V_{\text{pp}}$ equals 0 V, the intensity of the incident laser light is constant and the scattered light is pseudothermal light \cite{martienssen}. As the $V_{\text{pp}}$ increases, the degree of second-order coherence increases, too. In order to study in detail how the degree of second-order coherence of superbunching pseudothermal light behaves as the applied voltage increases, we measured the output intensity of D$_I$ and $g^{(2)}_\text{s}(0)$ by increasing $V_{\text{pp}}$ with a step of 0.5 V from 0 V to 8 V. Figure \ref{g2-C} shows that $g^{(2)}_\text{s}(0)$ of the superbunching pseudothermal light and the coefficient, $C$, of the incident laser light intensity increase as $V_{\text{pp}}$ increases. When $V_{\text{pp}}$ is less than 2.5 V, the intensity of the incident laser light is still modulated. However, the modulation is not large enough to ensure that  $g^{(2)}_\text{s}(0)$ becomes larger than 2. 

\begin{figure}[htbp]
\centering
\includegraphics[width=95mm]{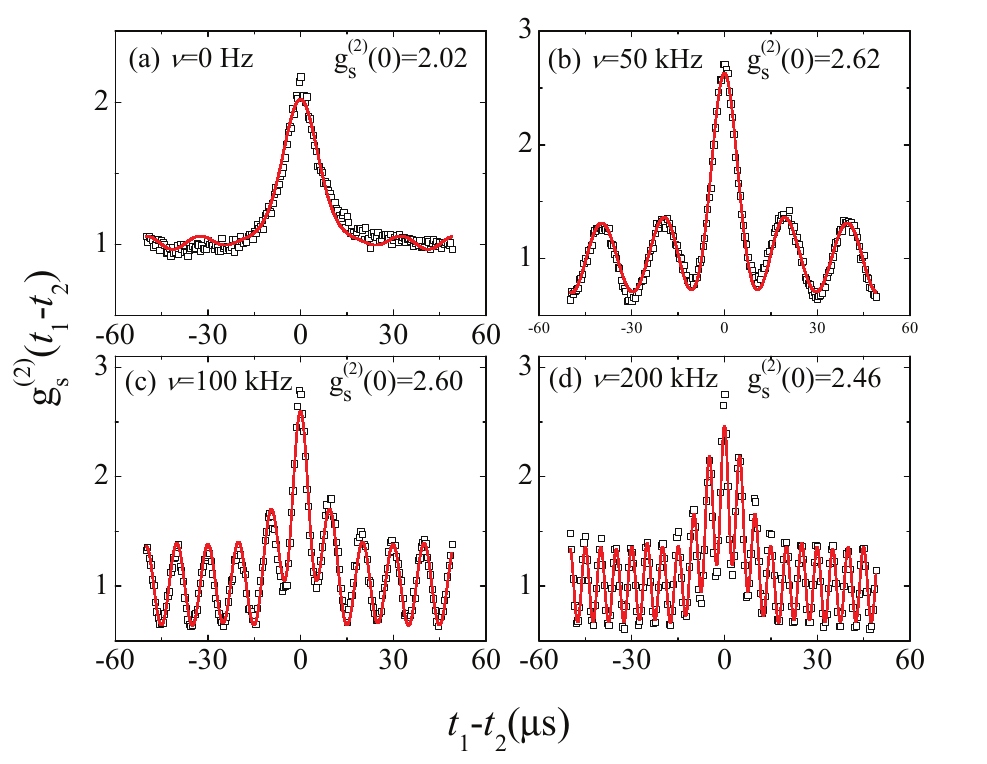}
\caption{The second-order temporal coherence function of superbunching pseudothermal light with sinusoidal voltage signals applied on EOM by varying frequency when $V_{\text{pp}}$ is set to be 8 V. $\nu$ is the frequency of sinusoidal voltage signal.}\label{frequency-s}
\end{figure}

Figure \ref{frequency-s} shows the measured second-order coherence function by varying the frequency of the sinusoidal voltage signals when  $V_{\text{pp}}$ equals 8 V. $\nu$ is the frequency of sinusoidal voltage signal. The means of the symbols are the same as the ones in Fig. \ref{voltage} and Eq. (\ref{G2-s}) is employed for the theoretical fitting. The reason why $g^{(2)}_\text{s}(0)$ equals 2.02 when the frequency equals 0 Hz is that the frequency of the applied sinusoidal voltage signals is actually 1 $\mu$Hz instead of 0 Hz in the experiments. Hence the intensity of the incident laser light in this case is not a constant. As the frequency of the sinusoidal voltage signals increases, the period of measured second-order coherence function decreases. The observed $g^{(2)}_\text{s}(t_1-t_2)$ in Fig. \ref{frequency-s} is a product of the correlations produced by light passing through EOM and RG, which is consistent with the predictions in Eq. (\ref{G2-s}).

\begin{figure}[htbp]
\centering
\includegraphics[width=95mm]{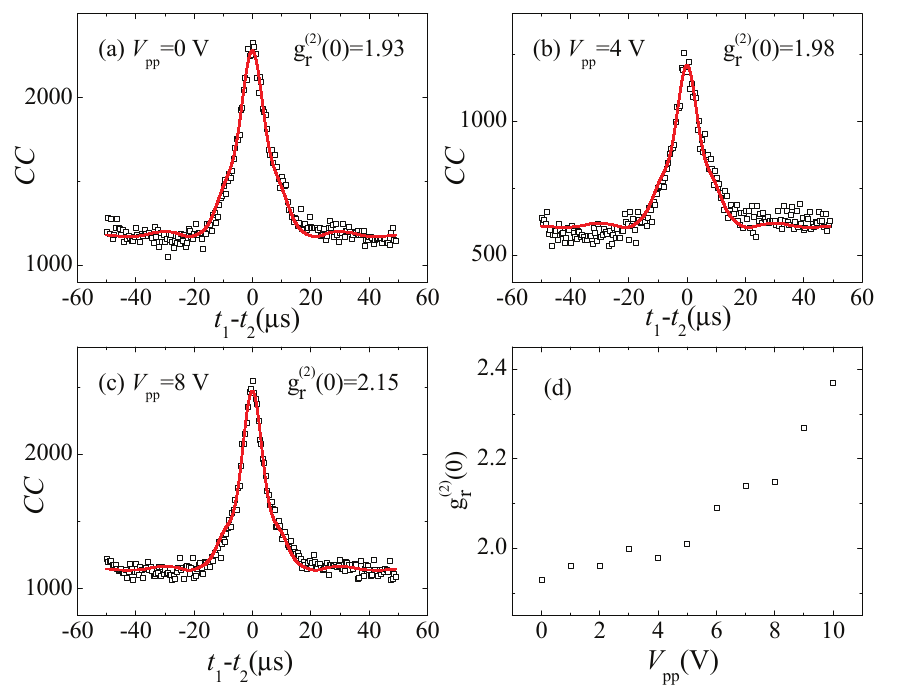}
\caption{The second-order temporal coherence function of superbunching pseudothermal light with random voltage signals applied on EOM by varying $V_{\text{pp}}$. The frequency of the white noise is fixed to be 200 Hz. $g^{(2)}_\text{r}(0)$ is the degree of second-order coherence with random white noise voltage signals.}\label{voltage-r}
\end{figure}

Comparing to the second-order temporal coherence function of pseudothermal light \cite{martienssen}, there is a sinusoidal modulation in Figs. \ref{voltage} and \ref{frequency-s} instead of being a constant. This problem is caused by the periodicity of sinusoidal wave, which can be solved by applying random voltage signals. Figure \ref{voltage-r} shows that the measured second-order temporal coherence function by applying white noise voltage generated by SG and varying the peak-to-peak voltage, $V_{\text{pp}}$. The frequency of the white noise is 200 Hz in the measurements. The empty squares in Figs. \ref{voltage-r}(a) - \ref{voltage-r}(c) are the measured second-order temporal coherence functions and red curves are theoretical fittings of the data by employing Eq. (\ref{G2-r}). The empty squares in Fig. \ref{voltage-r}(d) are the calculated $g^{(2)}_\text{r}(0)$ by varying $V_{\text{pp}}$ when the frequency is fixed to be 200 Hz. The degree of second-order coherence increases as $V_{\text{pp}}$ increases. The maximal value of  $g^{(2)}_\text{r}(0)$ in this case is predicted to be 4 based on Eq. (\ref{G2-r}). In our experiments, we have observed $g^{(2)}_\text{r}(0)$ equaling 2.73, which is less than 4. However, our experiments confirmed that two-photon superbunching can be observed by modulating the intensity of the incident laser light.

\begin{figure}[htbp]
\centering
\includegraphics[width=95mm]{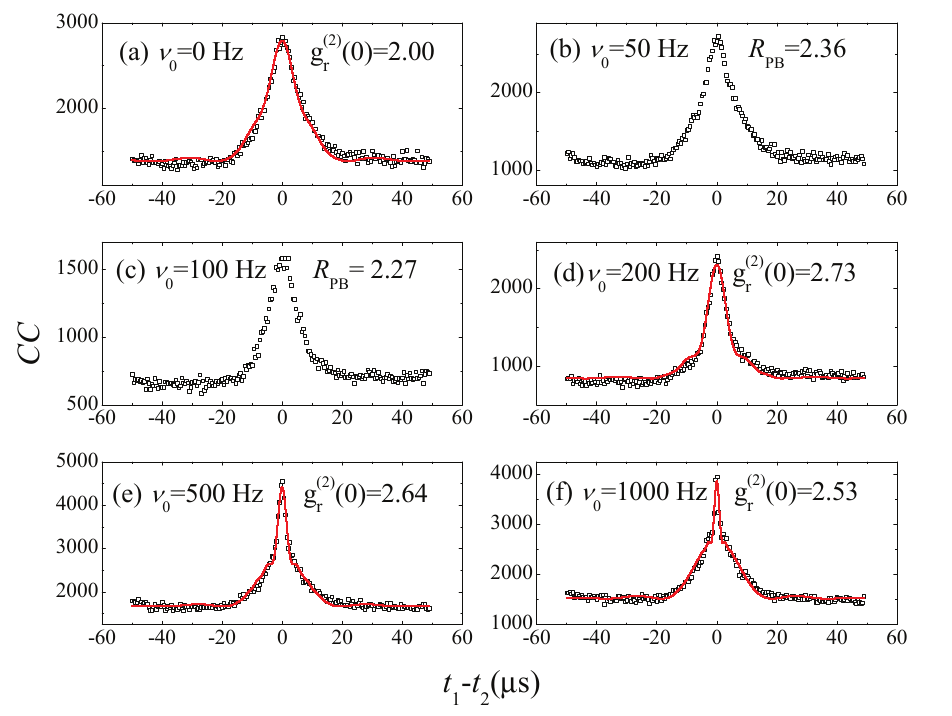}
\caption{The second-order temporal coherence function of superbunching pseudothermal light with random voltage applied on EOM by varying the frequency. The applied voltage, $V_{\text{pp}}$, is fixed to be 10 V. $CC$ is two-photon coincidence counts and $R_{\text{PB}}$ is the ratio between the measured peak and background $CC$.}\label{frequency-r}
\end{figure}

We also measured the second-order coherence function of superbunching pseudothermal light by varying the frequency of the applied white noise voltage signals when $V_{\text{pp}}$ is fixed to be 10 V.  $CC$ is two-photon coincidence counts and $R_{\text{PB}}$ is the ratio between the peak and background of the measured $CC$. The empty squares are measured results and red curves in Figs. \ref{frequency-r}(a), \ref{frequency-r}(d) - \ref{frequency-r}(f) are theoretical fittings of the data by employing Eq. (\ref{G2-r}). $R_{\text{PB}}$ in Figs. \ref{frequency-r}(b) and \ref{frequency-r}(c) does not equal $g^{(2)}(0)$ since the background $CC$ is not really the constant background $CC$. The correlation time caused by EOM in these two cases are much longer than the measurement time window, 100 $\mu$s. Hence it is impossible to obtain the constant background $CC$ in these two conditions. On the other hand, the constant background $CC$ is obtained in Fig. \ref{frequency-r}(f), in which the correlation time caused by EOM is much shorter than the one caused by RG. Hence it can be inferred that the constant background $CC$ is also obtained in Figs. \ref{frequency-r}(d) and \ref{frequency-r}(e) by noticing that the calculated values of $g^{(2)}_\text{r}(0)$ in the last three sub-figures are at the same level.

\section{Discussions}

In the above section, we have experimentally confirmed that two-photon superbunching can be obtained by varying the intensity of the incident laser light before RG. There are two interesting points worth noticing. The first one is that the measured degree of second-order coherence of superbunching pseudothermal light increases as $V_{\text{pp}}$ increases while it does not change when the frequency of the applied voltage signals varies. The conclusion holds for both sinusoidal and white noise voltage signals. This phenomenon can be easily understood for sinusoidal voltage signal by calculating the degree of second-order coherence with the help of Eq. (\ref{G2-s}). The degree of second-order coherence of superbunching pseudothermal light with sinusoidal voltage signal can be expressed as
\begin{eqnarray}\label{g0-s}
g^{(2)}_s(0)=\frac{(1+2C)\times 2}{(1+C)\times 1}=2+\frac{2C}{1+C},
\end{eqnarray}
in which $1+2C$ and $2$ in the numerator correspond to the maximal correlations caused by EOM and RG, respectively. $1+C$ and $1$ in the denominator correspond the constant background of the correlations introduced by EOM and RG, respectively. $g^{(2)}_s(0)$ equals 2 when $C$ equals 0, which is the typical value of thermal or pseudothermal light \cite{martienssen}. When  $C$ equals 1, $g^{(2)}_s(0)$ will reach its maximum, 3. Based on the results in Fig. \ref{g2-C}, $C$ increases from 0 to 0.94 as $V_{\text{pp}}$  increases from 0 V to 7.5 V. It is obvious that $g^{(2)}_s(0)$ will increase as $V_{\text{pp}}$ increases. There is no frequency in Eq. (\ref{g0-s}). Hence the change of frequency will not influence the value of $g^{(2)}_s(0)$.

Things become different for white noise voltage signals. The calculated degree of second-order coherence function, $g^{(2)}_r(0)$, equals 4 based on Eq. (\ref{G2-r}), which is independent of the frequency and voltage of white noise signals. The results in Fig. \ref{frequency-r} confirm that $g^{(2)}_r(0)$ is independent of frequency. However, the results in Fig. \ref{voltage-r} indicates that $g^{(2)}_r(0)$ increases as $V_{\text{pp}}$ increases. The reasons why the theoretical prediction in Eq. (\ref{G2-r}) is different from the experimental results are as follows. In the theoretical calculations, we employ perfect white noise model, which assumed that the intensity of the modulated laser light can be any value required by the distribution. While in the experiments, it is impossible to have intensity larger than the maximal value of the incident laser light. Any value exceeding the maximum will be treated as maximum in the experiments, which will decrease the variance of the intensity fluctuations. Another important reason is that the signal generator (SG) employed in our experiments only has 8 bits resolution, which means there are only 256 different values of intensity possible in our experiments. These two effects are the main reasons why the measured results are different from the theoretical predictions when white noise voltage signals are applied on the EOM, which can also be employed to interpret why the measured $g^{(2)}_r(0)$ is less than 4.

The other point worthy of noticing is that the frequency of sinusoidal voltage signal is different from the one of white noise voltage signal when the correlation time caused by EOM are at the same level. For instance, comparing the results in Figs. \ref{voltage}(d)  and \ref{frequency-r}(d), the width of the correlation time with 50 kHz sinusoidal wave signal is close to the one with 200 Hz white noise signal. The reason is that the definition of white noise frequency is different from the definition of the sinusoidal wave frequency. For instance, if the frequency of a sinusoidal wave is $\nu$, it means that the sinusoidal wave only have one frequency component, $\nu$. On the other hand, if the frequency of the white noise is $\nu$, it means that the white noise signal have frequency components extending from 0 to $\nu$. This can be easily understood by comparing the second-order temporal coherence functions in Eqs. (\ref{G2-s}) and (\ref{G2-r}).  

As mentioned before, the observed two-photon superbunching effect in the scheme shown in Fig. \ref{setup} can be divided into two parts. One part is caused by the scattering of laser light from rotating groundglass, which is expressed by $ [1+ \text{sinc}^2\frac{\Delta \omega (t_1-t_2)}{2}]$ in Eq. (\ref{G2-5}). The other part is caused by the intensity fluctuation of the laser light before RG, which is expressed by $\Gamma(t_1-t_2)$ in Eq. (\ref{G2-5}).  The correlation of the first part, \textit{i.e.}, pseudothermal light, have been studied extensively for decades \cite{goodman-speckle}, which has also been widely applied in ghost imaging \cite{valencia,ferri,crosby}, multi-photon interference \cite{scarcelli,chen}, non-invasive imaging \cite{bertolotti,katz}, and so on. However, there are only several studies about the coherence properties of superbunching pseudothermal light \cite{zhou, liu-5} and its application in ghost imaging \cite{liu-6}. The coherence properties of superbunching pseudothermal light are worthy of studying and the applications of superbunching pseudothermal light are also worthy of exploring. The proposed superbunching pseudothermal light source, which can be easily implemented in most optics laboratories, may be beneficial for the community to study the coherence properties and find possible applications of superbunching pseudothermal light.
 
\section{Conclusions}

By varying the intensity of the incident laser light before rotating groundglass, the coherence properties of the scattered light can be tuned.  With suitable intensity fluctuation, superbunching pseudothermal light can be obtained. We have theoretically and experimentally proved that the degree of second-order coherence of superbunching pseudothermal light can be larger than 2 with sinusoidal wave and white noise modulations. The degree of second-order coherence of superbunching pseudothermal light can be further tuned by changing the intensity fluctuation models.  It is expected that three-photon and multi-photon superbunching can also be observed with the proposed superbunching pseudothermal light \cite{liu-5}. This new type of light is expected be helpful to understand the physics of two-photon bunching and superbunching, which is important to understand the difference between quantum and classical interpretations of the second- and higher-order interference of light.

\section*{Acknowledgement}
This project is supported by National Natural Science Foundation of China (NSFC) (Grant No.11404255), 111 Project of China (Grant No.B14040), and Fundamental Research Funds for the Central Universities.

\end{document}